\documentclass[aps,prd,twocolumn,showpacs,preprintnumbers,nofootinbib]{revtex4-1}

\usepackage{graphicx}
\usepackage{dcolumn}
\usepackage{bm}
\usepackage[normalem]{ulem}

\usepackage{amsmath}
\usepackage{epstopdf}
\usepackage{amsmath,amssymb}
\usepackage[normalem]{ulem}

\newcommand{\be}{\begin{equation}}
\newcommand{\ee}{\end{equation}}
\newcommand{\ba}{\begin{eqnarray}}
\newcommand{\ea}{\end{eqnarray}}

\newcommand{\beq}{\begin{eqnarray}}
\newcommand{\eeq}{\end{eqnarray}}

\def\spose#1{\hbox to 0pt{#1\hss}}
\def\ltapprox{\mathrel{\spose{\lower 3pt\hbox{$\mathchar"218$}}
 \raise 2.0pt\hbox{$\mathchar"13C$}}}

\begin{document}

\title{
The chiral phase transition in two-flavor QCD from
imaginary chemical potential
}

\author{Claudio Bonati}
\email{bonati@df.unipi.it}
\affiliation{Dipartimento di Fisica dell'Universit\`a
di Pisa and INFN - Sezione di Pisa,\\ 
Largo Pontecorvo 3, I-56127 Pisa, Italy}

\author{Philippe de Forcrand}
\email{forcrand@phys.ethz.ch}
\affiliation{
Institute for Theoretical Physics, ETH Z\"urich, 
CH-8093 Z\"urich, Switzerland and\\
CERN, Physics Department, TH Unit, CH-1211 Geneva 23, Switzerland}

\author{Massimo D'Elia}
\email{delia@df.unipi.it}
\affiliation{
Dipartimento di Fisica dell'Universit\`a
di Pisa and INFN - Sezione di Pisa,\\ 
Largo Pontecorvo 3, I-56127 Pisa, Italy}

\author{Owe Philipsen}
\email{philipsen@th.physik.uni-frankfurt.de}
\affiliation{
Institut f\"ur Theoretische Physik, Goethe-Universit\"at Frankfurt, \\ 
60438 Frankfurt am Main, Germany}

\author{Francesco Sanfilippo}
\email{f.sanfilippo@soton.ac.uk}
\affiliation{School of physics and astronomy, University of Southampton, 
SO17 1BJ Southampton, United Kindgdom}

\preprint{CERN-PH-TH-2014-158}
\preprint{IFUP-TH/2014-8}
 
\date{\today}

\begin{abstract}
We investigate the order of the finite temperature chiral symmetry restoration
transition for QCD with two massless fermions, by using a novel method, based
on simulating imaginary values of the quark chemical potential
$\mu=i\mu_i,\mu_i\in\mathbb{R}$.
Our method exploits the fact that, for low enough quark mass $m$ and large
enough chemical potential $\mu_i$, the chiral transition is decidedly first
order, then turning into crossover at a critical mass $m_c(\mu)$.  It is thus
possible to determine the critical line in the $m - \mu^2$ plane, which can be
safely extrapolated to the chiral limit by taking advantage of the known
tricritical indices governing its shape.
We test this method with standard staggered fermions and the result of our
simulations is that $m_c(\mu=0)$ is positive, so that the phase transition at
zero density is definitely first order in the chiral limit, on our coarse
$N_t=4$ lattices with $a\simeq 0.3\,\mathrm{fm}$.

\end{abstract}

\maketitle

\section{Introduction}\label{sec:intro}

Quantum Chromodynamics is known to undergo a finite temperature ``transition''
at which both partial deconfinement and partial chiral symmetry restoration
take place.  Because of its relevance to heavy ion collisions, the early
universe and astrophysics, the nature of this QCD transition is the subject of
intense numerical studies. Present results show the absence of a true
non-analytic phase transition in the case of physical quark
masses~\cite{nature, tchot, tchot2}, so that deconfinement would just
correspond to a rapid analytic change of the effective degrees of freedom
(crossover). This is consistent with the fact that the global symmetries of QCD
associated with the transition are only exact in the limits of infinite quark
masses (center symmetry) or of massless quarks (chiral symmetry): only in these
limits is a true phase transition expected a priori, representing a change in
the realization of the corresponding symmetry.

\begin{figure}[t]
\begin{center}
\includegraphics*[width=0.35\textwidth]{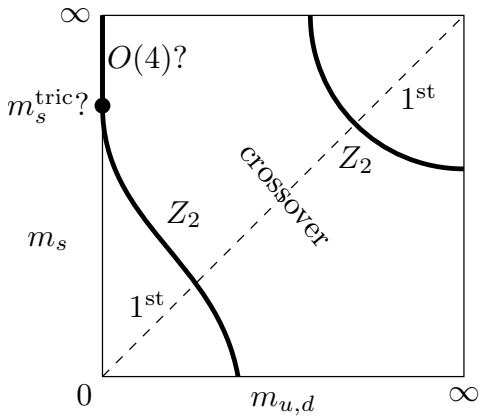}
\end{center}
\caption{Schematic behavior of the order of the finite temperature phase
transition for $N_f=2+1$ QCD as a function of the up/down and strange quark
masses $(m_{u,d},m_s)$, at zero baryon chemical potential. $O(4)$ and $Z_2$
denote transitions of the second order in the universality class of
the $3d$ $O(4)$ and Ising models, respectively.}
\label{fig:columbia}
\end{figure}

However, a full understanding of the QCD phase diagram, and in particular of
the interplay between center symmetry and chiral symmetry, requires to study
the phase transition as a function of the quark masses.  The current state of
knowledge at zero baryon density is summarized in Fig.~\ref{fig:columbia}.  For
three degenerate flavors, QCD displays a first order phase transition in the
limits of zero and infinite quark masses, associated with the breaking of the
chiral and center symmetries, respectively.  For finite quark masses, instead,
these symmetries are broken explicitly; the associated first order transitions
weaken away from these limits, until they disappear at critical points,
belonging to the Ising ($Z_2$) universality class.  The chiral
\cite{chcrit, mu4} and deconfinement \cite{deccrit, saito, saito1} critical
lines are known on coarse $N_t=4$ lattices, based on simulations with staggered
and Wilson fermions, respectively.

\begin{figure}[b] 
\begin{center}
\includegraphics*[width=0.35\textwidth]{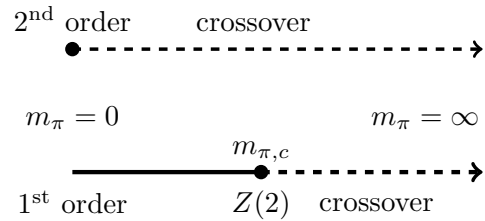}
\end{center}
\caption{Possible scenarios for the chiral phase transition 
as function of the pion mass.}
\label{fig:nf2}
\end{figure}

A still open issue is the order of the transition
in the limit of two massless flavors (upper left corner of
Fig.~\ref{fig:columbia}). This is in fact a longstanding problem, with a
history of conflicting lattice results between Wilson~\cite{Tsukuba,cp-pacs}
and staggered, and even within staggered
\cite{fuku1,fuku2,columbia,karsch1,karsch2,jlqcd,milc,DiG,Unger} fermions.  The
possibilities for continuum QCD have been discussed systematically in
Refs.~\cite{pw,vicari,vicari2}.  The classical action has a global
$SU(2)_R\times SU(2)_L\times U_A(1)$ chiral symmetry, with the $U_A(1)$
undergoing anomalous breaking and expected to be effectively restored at
high temperature.  Since the phase transition is associated with a change in
the realization of an exact symmetry, an analytic crossover is ruled out,
leaving room for either a first order or second order transition.  In the case
of a second order transition, the corresponding universality class is fixed by
the critical behavior of the associated effective chiral model: that can be
$SU(2)_L\times SU(2)_R\simeq O(4)$, or $U(2)_R\times
U(2)_L/U(2)_V$~\cite{vicari,vicari2}, depending on the strength of the axial
anomaly at the transition temperature.  A first order transition,
instead, can be present independently of the critical behavior of the chiral
effective model, though it is generally believed that a vanishing strength of
the axial anomaly could make it more likely~\cite{pw,vicari,mehta,vicari2}. 
In fact, the presence or absence of a first order transition is something which
depends on the interplay among the degrees of freedom of the strong
interactions which are effective around the transition, and can be decided only
on the basis of direct numerical simulations of QCD.  Once finite quark masses
are switched on, a second order transition disappears immediately, while a
first order transition gets gradually weakened, until it disappears at a $Z_2$
critical point.  The two possible cases are sketched in Fig.~\ref{fig:nf2}. 

Unfortunately, distinguishing between these two scenarios (first or second
order transition) by lattice simulations is notoriously difficult.  Since
massless fermions cannot be directly simulated, the standard strategy is to
perform simulations at a number of finite quark masses, trying to approach the
chiral limit. If the transition is second order, then one should observe the
scaling relations expected near the critical point. If the transition is first
order, one possibility is to directly detect the first order line at finite quark
mass, by observing the associated discontinuities and metastabilities.
Alternatively, if the critical quark mass $m_c$ at which the first order
disappears is too small to be probed directly, one should look for the
$Z_2$ critical behavior near the critical endpoint.

Various studies have found no signs of metatabilities in the range of masses
explored up to now, thus failing a direct detection of first order. On the
other hand, as we discuss in the following, the analysis of the critical
behavior is still not conclusive.  Most scaling relations are given in terms of
the symmetry breaking parameter, i.e. the quark mass $m$. Examples are given
by the divergent part of the susceptibility of the order parameter (i.e. the
chiral susceptibility): 
\begin{equation}\label{eq:chichiralfit} 
\chi (m) \sim B \, m^{(1-\delta)/\delta}\; 
\end{equation} 
or, alternatively, by the pseudo-critical temperature as a function of the
quark mass: 
\begin{equation} \label{eq:tcchiralfit} 
T_c(m) = T_c(0) + A \, m^{1/(\beta\delta)}\; .
\end{equation}

Unfortunately, the critical behavior turns out to be 
very similar for the
various universality classes which could be relevant.  For instance, the
behavior of the chiral susceptibility turns out to be consistent with the
$O(4)$ scenario ($1 - 1/\delta \simeq 0.79$, see \cite{O4a, O4b}), but
it is hardly possible to distinguish it from that predicted by the
$U(2)_R\times U(2)_L/U(2)_V$ universality class (whose critical indices
are almost identical to that of the $O(4)$ case, see \cite{vicari2}), from
other $O(N)$ universality classes (\emph{e.g} $O(2)$, see \cite{O2}),
or even from the $Z_2$ critical behavior associated with a 
critical endpoint mass $m_{c}$ close to zero ($1 -1/\delta \simeq
0.79$ for $Z_2$, see \cite{Z2}).

The same is true for the behavior of the pseudocritical temperature, as noted
both for staggered~\cite{DiG} and Wilson~\cite{tmft} fermions, since the
critical exponents are very close, e.g.  $1/(\beta\delta)\simeq 0.537,0.639$ for
$O(4),Z_2$, respectively (see \cite{O4a, O4b, Z2}), so that there is no way of
distinguishing the correct scenario within the range of currently explorable
quark masses. A more sensitive quantity would be the specific heat~\cite{DiG},
whose critical exponent is $\alpha \simeq -0.24, 0.11, 1$ for $O(4)$, $Z_2$ and first order
respectively (see \cite{O4a, O4b, Z2}), but unfortunately determining $\alpha$ by lattice
simulations is made difficult by the presence of large non-singular
contributions.

In the present paper, we propose an alternative approach to solve this problem,
which is based on the investigation of the phase diagram in the presence of a
purely imaginary quark chemical potential $\mu=i\mu_i$.  This is usually
exploited to partially circumvent, via analytic continuation, the sign problem
which is met in QCD at finite baryon density. However, the phase diagram at
imaginary $\mu$ turns out to be quite interesting by itself and, by exploiting
its peculiar features, we can extract useful information for the problem at
hand.

Our approach is based on the clear evidence of the first order nature of the
transition at large enough imaginary chemical potentials and low quark masses.
The nature of the transition in the chiral limit can be
explored by determining the maximal quark mass necessary to keep the transition
first order while driving the chemical potential to zero.  While answering this
question might appear as difficult as (or even harder than) 
the original problem, the issue is greatly simplified by taking advantage of
the universal tricritical scaling, which imposes strong constraints on the
boundary of the first order region, as discussed below.

The paper is organized as follows.  In Sec.~\ref{sec:phase_diag} we discuss the
general form of the QCD phase diagram at imaginary chemical potential.  Some of
its properties are used in Sec.~\ref{sec:nf2} to propose our new method to assess
the problem of the determination of the order of the $N_f=2$ chiral transition
at $\mu=0$. In Sec.~\ref{sec:numerical} we describe our numerical setting and
the results obtained by using the introduced method. Finally in
Sec.~\ref{sec:conclusions} we present our conclusions and perspectives for
future studies. Preliminary results have been presented in \cite{pos}.

\section{Phase diagram for imaginary chemical potential}\label{sec:phase_diag}

\begin{figure}[t]
\begin{center}
\includegraphics*[width=0.35\textwidth]{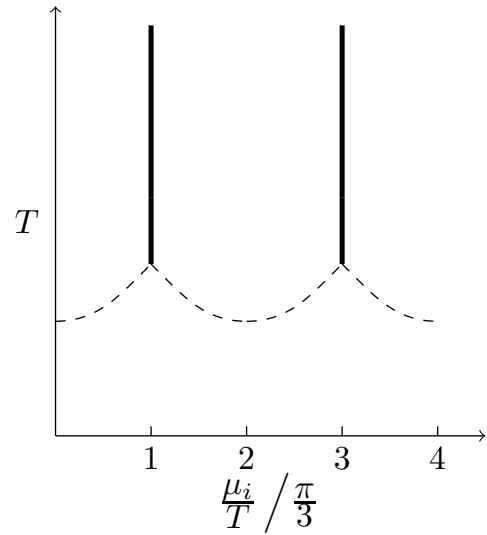}
\end{center}
\caption{Generic phase diagram as a function of imaginary chemical potential 
and temperature. Solid lines are first order Roberge-Weiss transitions. 
The behavior along dashed lines depends on the number of flavors and 
the quark masses.} 
\label{fig:rw}
\end{figure}

When the chemical potential is considered as a generic complex parameter
entering the definition of the partition function $Z$, the two following exact
symmetries emerge (\cite{RW}) 
\begin{equation}
Z(\mu)=Z(-\mu), 
\quad 
Z\left(\frac{\mu}{T}\right) = Z\left(\frac{\mu}{T} + i\frac{2\pi n}{3}\right)
\quad
n\in\mathbb{Z}\ ,
\end{equation}
which imply reflection symmetry in the imaginary $\mu$ direction about the
``Roberge-Weiss'' values $\mu = i\pi T/3 (2n+1)$.  Along these lines a $Z_2$
exact symmetry is present \cite{RW}, which can be explicitly realized or
spontaneously broken, different sectors corresponding to different preferred
orientations of the Polyakov loop. Transitions between neighboring sectors are
of first order for high $T$ \cite{RW} and analytic crossovers for low $T$
\cite{fp1,el1,ddl}.  
The first order transition lines at $\mu = i\pi T/3 (2n+1)$ may end with a
second order critical point or with a triple point, branching off into two
first order lines. One of these may continue to zero and real chemical
potential and represents the analytic continuation of the chiral or
deconfinement transition. For the deconfinement transition this has been
explicitly demonstrated in \cite{deccrit}, for the chiral transition it is
demonstrated in the present work.  Which of these possibilities occurs depends
on the number of quark flavors and on their masses.  From these considerations, it
follows that the phase diagram in the $T-\mu_i$ plane is of the form indicated
in Fig.~\ref{fig:rw}. 

\begin{figure}[b]
\begin{center}
\includegraphics*[width=0.35\textwidth]{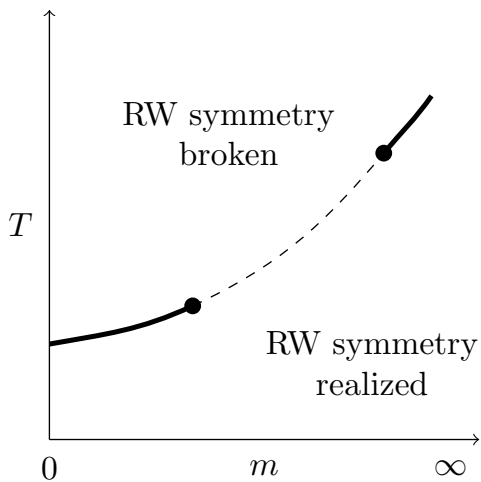}
\end{center}
\caption{Schematic picture of the phase diagram for $N_f=2$ and $N_f=3$ at the
Roberge-Weiss point $\mu_i/T=\pi/3$. Solid lines denote first order
transitions, the dashed line indicates a second order transition of the $3d$
Ising universality class and the two dots represent the tricritical points, at
which the order of the transition changes.} 
\label{fig:t_m}
\end{figure}

Recent numerical studies have shown that the triple point case is found for
heavy and light quark masses, while for intermediate masses one finds a
second order endpoint (this happens both for $N_f=2$ \cite{CGFMRW, PP_wilson}
and for $N_f=3$ \cite{OPRW}).  As a function of the (increasing) quark masses,
the finite temperature transition for the theory with fixed $\mu/T
= i\pi/3$ thus switches from being first order to second order and then to
first order again.  The points at which the change of the order takes place are
tricritical points and a sketch of this dependence of the order of the
transition on the quark masses is depicted in Fig.~\ref{fig:t_m}. Results have
first been obtained within a staggered fermion formulation of QCD, but
efforts have been undertaken to confirm their universality also within a Wilson
fermion approach~\cite{nakamura, wumeng, PP_wilson, alexandru} leading to the
same qualitative phase diagram~\cite{PP_wilson}. The Roberge-Weiss endpoint
transition, or variants of it, has been also studied in many other different
contexts and QCD-like theories~\cite{lucini07, cea, kouno, holorw, holorw2,
Braun, morita, weise, kp13, rw-2color, pagura, buballa}.

\begin{figure}[b]
\begin{center}
\includegraphics*[width=0.35\textwidth]{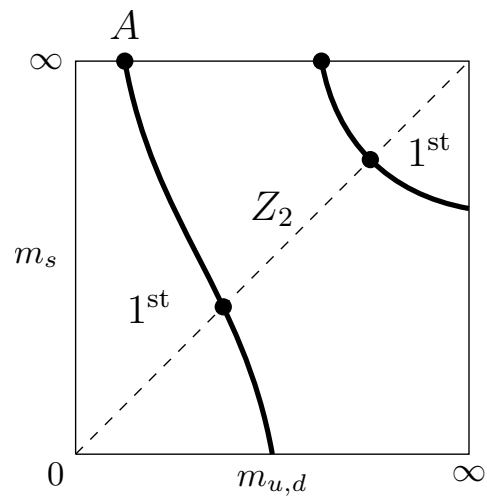}
\end{center}
\caption{The behavior of the order of the finite temperature transition in
$N_f=2+1$ QCD at the Roberge-Weiss point $\mu_i/T=\pi/3$. The regions
denoted by $1^{\mathrm{st}}$ and $Z_2$ represent respectively regions where
first order transitions and second order transitions of the universality class
of the $3d$ Ising model take place.  The solid lines are lines of tricritical
points and the dashed line is the line $m_s=m_{u,d}$ corresponding to $N_f=3$
QCD. The dots represent the tricritical points found in previous $N_f=2$
(\cite{CGFMRW, PP_wilson}) and $N_f=3$ (\cite{OPRW}) simulations.}
\label{fig:columbia_rw}
\end{figure}

Let us now consider the analogue of the phase diagram of
Fig.~\ref{fig:columbia}, but at the Roberge-Weiss value of the imaginary
chemical potential: $\mu/T = i\pi/3$.  It is important to stress that in this
case the finite temperature transition corresponds to the breaking of a $Z_2$
symmetry, thus the possibility of an analytic crossover can be excluded
\emph{a priori}.  The simplest expectation is that the
tricritical points observed for $N_f=2$ and $N_f=3$ are connected to each other
in the $(m_{u,d},m_s)$ quark mass plane, the resulting phase diagram being
depicted in Fig.~\ref{fig:columbia_rw}: two tricritical lines separate regions
of first order and of second order transitions.  Note that the assumption of
continuity of the tricritical lines can be checked directly by numerical
simulations, since there is no sign problem for imaginary $\mu$.

We now have to connect the two phase diagrams in Fig.~\ref{fig:columbia} and
Fig.~\ref{fig:columbia_rw} to obtain the complete phase diagram in the three
dimensional space $(m_{u,d},m_s,\mu^2)$.  When there are no symmetry constraints
second order transitions are expected to be washed out by perturbations, while
first order transitions are robust.  As a consequence, we can expect the $Z_2$
region at $\mu/T=i\pi/3$ to be connected with the crossover region at $\mu =0$,
while the two regions of first order transitions at small and large masses
become nontrivial three dimensional volumes, 
which extend from the $\mu=0$ plane
to the Roberge-Weiss $\mu^2$ value. 

\begin{figure}[bt]
\begin{center}
\includegraphics*[width=0.4\textwidth]{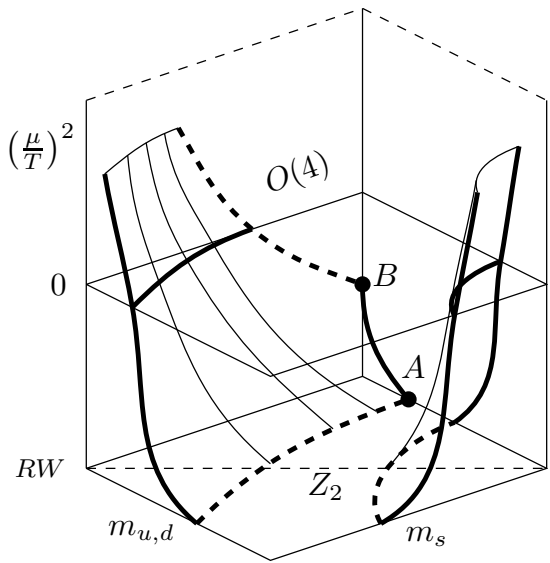}
\end{center}
\caption{The same as in Fig.~\ref{fig:columbia}, with the quark chemical
potential $\mu$ as an additional parameter.  The critical boundary lines sweep
out surfaces as $\mu$ is turned on.  At $\mu/T=i\pi/3$ (value denoted by $RW$),
as well as at $m_{u,d} = 0$, the critical surfaces terminate in tricritical
lines (represented by bold dashed lines), which fix the curvature of the surfaces
through tricritical scaling.}
\label{fig:columbia_3d}
\end{figure}

We thus expect a phase diagram of the type depicted in
Fig.~\ref{fig:columbia_3d}, in which the three dimensional first order regions
are bounded by surfaces of second order phase transitions in the universality
class of the $3d$ Ising model. 
Besides the first order regions and the critical surfaces, one or possibly 
two planar
regions of second order transitions are present: 
one is in the bottom plane of
Fig.~\ref{fig:columbia_3d} (and corresponds to the second order region in
Fig.~\ref{fig:columbia_rw}), the possible other is in the plane $m_{u,d}=0$, 
where the
relevant symmetry 
is chiral symmetry.

\section{The $N_f=2$ chiral transition and tricritical scaling}\label{sec:nf2}

\begin{figure}[t]
\begin{center}
\includegraphics*[width=0.35\textwidth]{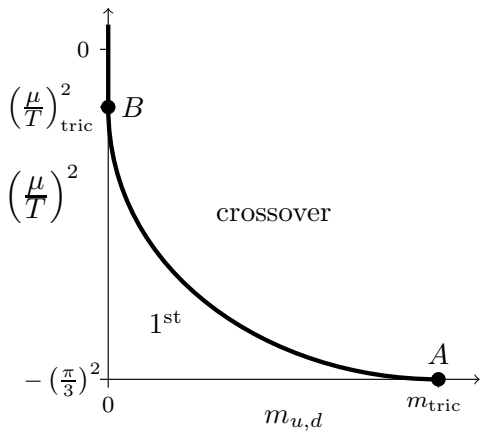}
\caption{The $N_f=2$ (i.e. $m_s=\infty$) section of Fig.~\ref{fig:columbia_3d}.
The two tricritical points at $m_{u,d}=0$ and $\mu=i\pi T/3$ are connected by a
$Z_2$ critical line. As in Fig.~\ref{fig:columbia_3d} this picture is plotted
assuming the $N_f=2$ finite temperature transition to be second order in the
chiral limit.}
\label{fig:nf2_sec}
\end{center}
\end{figure}

Let us take a look at the $N_f=2$ section of the phase diagram in
Fig.~\ref{fig:columbia_3d}, i.e. at the plane $m_s=\infty$, and, in particular,
at its first order region present for small values of $m_{u,d}$, which is shown
in Fig.~\ref{fig:nf2_sec}. This picture (as well as Fig.~\ref{fig:columbia} and
Fig.~\ref{fig:columbia_3d}) is drawn by assuming the $N_f=2$ chiral transition
to be second order for $\mu=0$, so that two tricritical points are present: the
one at $(\mu/T)^2=-(\pi/3)^2$, $m_{u,d}=m_{\mathrm{tric}}$, which is labeled
``A'' and is just the upper point of the tricritical line shown in
Fig.~\ref{fig:columbia_rw}, and the one located at $m_{u,d}=0$,
$(\mu/T)^2=(\mu/T)^2_{\mathrm{tric}}<0$, which is connected to the tricritical
point present at $\mu=0$ (see Fig.~\ref{fig:columbia_3d}) and is labeled
``B''.

We can now come back to our original aim, that is the determination of the
order of the $N_f=2$ chiral transition at zero density: the two possible cases
correspond to $(\mu/T)^2_{\mathrm{tric}}<0$ (second order) and 
$(\mu/T)^2_{\mathrm{tric}}>0$ (first order).
Our strategy to attack this problem is thus:
\begin{enumerate}
\item map the line of $Z_2$ second order transition at $m_{u,d}>0$ and 
$(\mu/T)^2<0$,
\item extrapolate this line to $m_{u,d}=0$ and check if its intersection
with the axis happens at positive or negative values of $(\mu/T)^2$.
\end{enumerate}

The first task can be accomplished by using standard tools of finite size
scaling theory. The location of the transition line can be 
conveniently determined
by looking at the crossing points of
renormalization group invariant quantities measured for different values of the
parameters and on different volumes.

The extrapolation to the chiral limit of the curve obtained in this way appears
at a first sight a much more delicate point. However the critical line,
in the proximity of a tricritical point, has to follow a power law with known
critical indices (see \cite{LawrieSarbach}). For the case studied here the
scaling law of the critical line is given by
\begin{equation}\label{eq:tric_scaling}
m_{u,d}^{2/5}= C\left[\left(\frac{\mu}{T}\right)^2-
\left(\frac{\mu}{T}\right)^2_{\mathrm{tric}}\right]\ .
\end{equation}
The same constraint imposed by tricritical scaling has been used, e.g., 
in \cite{W92, RW93} to predict the shape of the $Z_2$ critical line in the 
proximity of the tricritical point in the phase diagram at $\mu=0$ 
(see Fig.~\ref{fig:columbia}).

By checking the agreement of our data with the scaling relation 
Eq.~\eqref{eq:tric_scaling} we can estimate the size of the scaling region
and, if we have enough data in the scaling region, we can use 
Eq.~\eqref{eq:tric_scaling} to safely estimate $C$ and $(\mu/T)^2_{\mathrm{tric}}$.

\section{Numerical results}\label{sec:numerical}

In this section we present the numerical results obtained using 
the method introduced in Sec.~\ref{sec:nf2}. The present study is
limited to the use of unimproved staggered fermions and lattices with 
temporal extent $N_t=4$, however the method is clearly general and 
it can be used to approach the continuum limit with no particular
technical difficulties. Numerical computations have been carried
out on standard CPU farms and on Graphics Processing Units (GPUs), 
exploiting the GPU code developed
in Ref.~\cite{gpucode}.

To map out the critical line shown in Fig.~\ref{fig:nf2_sec}, we used the 
crossing method for the fourth order cumulant
\begin{equation}\label{binder}
B_4(m,\mu)=\frac{\langle(\delta X)^4\rangle}
{\langle(\delta X)^2\rangle^2}\ ,
\end{equation}
where $\delta X= X-\langle X\rangle$ is the fluctuation of the variable of
interest (\cite{Binder81a, Binder81b}). Since we investigated the region of 
small masses, it was natural to take for $X$ the chiral condensate: 
$X = \bar{\psi}\psi$.  

For the application of this method various simulations were performed for
several values of the masses ($a m_{u,d}$), of the (imaginary) chemical
potential $\left( (a\mu)^2 \right)$ and of the bare coupling ($\beta$). For
each couple of $\left(am_{u,d}, (a\mu)^2\right)$ values we identified (by the
vanishing of the third moment, $\langle (\delta X)^3\rangle=0$) the
pseudo-critical value of the coupling, denoted by
$\beta_{pc}\left(am_{u,d},(a\mu)^2\right)$, that is the value which in the
thermodynamic limit converges to the critical value separating the low and
high-temperature phases. We then computed the value of the fourth order
cumulant 
\begin{equation}
B_4\left(am_{u,d},(a\mu)^2\right)\equiv B_4\left(am_{u,d},(a\mu)^2, \beta_{pc}\right)\ ,
\end{equation}
which is, in the thermodynamical limit, a discontinuous function of the
parameters.  Indeed $B_4=1$ if a first order transition is present (i.e. if the
distribution of $X$ is strongly peaked around two values), $B_4=3$ when there
is no transition (i.e. the $X$ distribution is Gaussian) while the value of the
$B_4$ parameter at a second order transition is universal and depends on the
scaling form of the $X$ distribution. In the particular case of the three
dimensional Ising model universality class this value is $B_4\approx 1.604$ (see
e.g. \cite{Z2}).

Discontinuities are smeared out in a finite volume and
$B_4\left(am_{u,d}, (a\mu)^2\right)$ passes continuously through the critical value. 
At fixed $am_{u,d}$ and in the neighborhood of the critical chemical potential,
denoted by $(a\mu_c)^2$, the function $B_4\left(am_{u,d},(a\mu)^2\right)$ 
can be expanded to leading order, obtaining (see e.g \cite{Binder81b})
\begin{equation}\label{bseries}
\begin{aligned}
& B_4\left(am_{u,d},(a\mu)^2\right)\approx \\
& \approx b_4^{(0)}(am_{u,d})+b_4^{(1)}L^{1/\nu}(am_{u,d}) \left((a\mu)^2-(a\mu_c)^2\right)\ ,
\end{aligned}
\end{equation}
with $b_4^{(0)}(am_{u,d})\approx 1.604$ and $\nu\approx 0.63$.

\begin{figure}[t]
\begin{center}
\includegraphics*[width=0.5\textwidth]{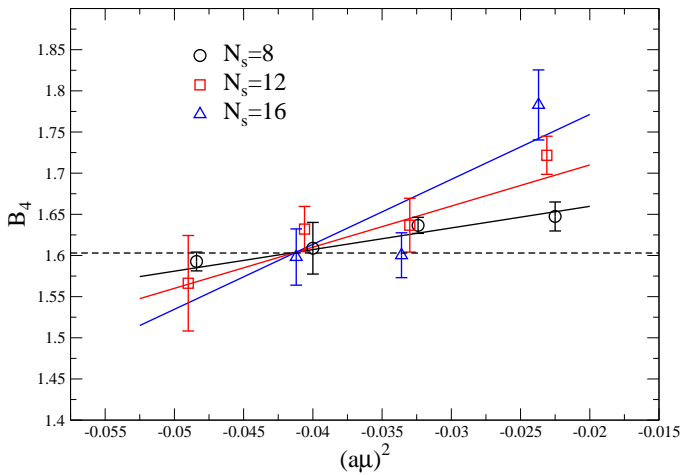}
\caption{Binder cumulant for fixed quark mass ($am_{u,d}=0.005$) as a 
function of imaginary chemical potential and volume. 
The intersection signals the critical point.}\label{fig:b4fit} 
\end{center}
\end{figure}

\begin{figure}[b]
\begin{center}
\includegraphics*[width=0.5\textwidth]{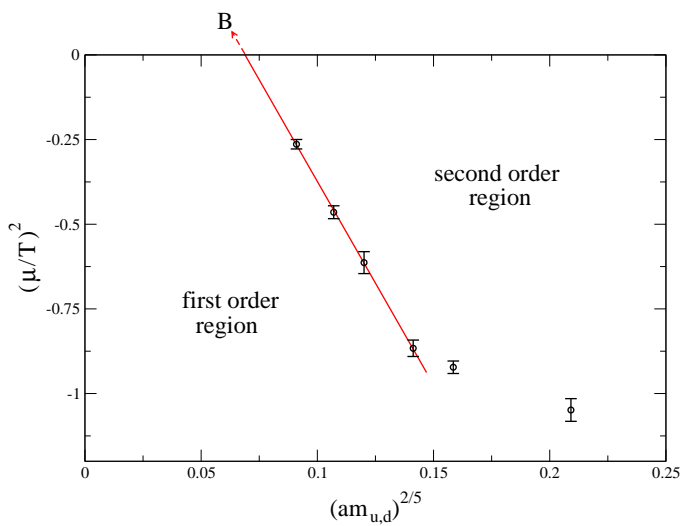}
\caption{Data corresponding to the calculated critical points, the line is
a fit according to tricritical scaling Eq.~\eqref{eq:tric_scaling}}
\label{fig:results} 
\end{center}
\end{figure}

By using this expression we can, for each value of the bare quark mass
$am_{u,d}$, find the corresponding critical value $(a\mu_c)^2$.  An example
from our data is shown in Fig.~\ref{fig:b4fit}, where (at fixed bare mass
$am_{u,d}=0.005$) we scanned in imaginary chemical potential using up to four
different volumes in order to identify the critical point and in all cases 
we reached lattice sizes such that $m_{\pi}L\gtrsim 3$ (in fact for all but the 
lightest mass used we arrived to $m_{\pi}L\gtrsim 4$).
The fit is performed simultaneously on all the data at different 
volumes and $b_4^{(0)}$ is fixed to its infinite volume limit.

This procedure was carried out for six different values of the quark mass 
and the results are shown Fig.~\ref{fig:results}.  
The quark mass axis is rescaled 
with the appropriate critical exponent in order to display the scaling and
extrapolation more clearly, as a straight line.  Four data points
accurately follow the tricritical scaling curve, which can then be used to
estimate the  position of the tricritical point ``B'' in the chiral
limit, for which we find the large positive value
\begin{equation}
\left( \frac{\mu}{T} \right)^2_{\mathrm{tric}}=0.85(5)\ .
\end{equation}
This definitely implies a first order behaviour for
the two-flavor chiral phase transition on $N_t=4$ lattices. A crude estimate
(obtained by using the interpolating formula for the masses of Ref.~\cite{BGKT})
puts the critical pion mass corresponding to the second order point at $\mu=0$
to $m_\pi^c\sim$ 60 MeV.

\section{Conclusions}\label{sec:conclusions}

We have presented a new approach for the determination of the order of
the chiral transition for $N_f = 2$ QCD, based on the investigation of the
phase diagram extended to imaginary chemical potential. In this approach, the
chiral limit extrapolation is controlled and constrained by scaling
considerations which follow from the universal behavior around a tricritical
point.  Present results show that, for QCD discretized on $N_t=4$ lattices with
standard staggered fermions, the transition is first order in the chiral limit.
This is consistent with some earlier lattice investigations~\cite{DiG}
and with expectations from the fate of the $U(1)_A$ anomaly using 
overlap fermions~\cite{overlap}.

It should be stressed that the explored $N_t=4$ lattice is quite coarse,
corresponding to $a\sim 0.3$ fm, and that results for $m_c(\mu)$  on
finer lattices are needed before a continuum limit can be taken.  For $\mu=0$ it is known
that the three-flavor chiral first order region in Fig.~\ref{fig:columbia}
(left) shrinks significantly on finer lattices \cite{kim} or with improved
actions \cite{ding}.  Therefore, the issue about the presence of a first order
chiral transition for $N_f = 2$ QCD in the continuum remains non-trivial.

We have shown that the proposed approach is able to provide definite answers
and constitutes a solid framework for future studies on the subject.

\section*{Acknowledgements}
O.P. is supported by the German BMBF, grant 06FY7100, and the Helmholtz
International Center for FAIR within the LOEWE program launched by the State of
Hesse.
F.S. has received funding from the
European Research Council under the European Community’s Seventh
Framework Programme (FP7/2007-2013) ERC grant agreement No 279757.

We thank the Scientific Computing Center at INFN-Pisa, INFN-Genoa, the HLRS
Stuttgart and the LOEWE-CSC at University of Frankfurt for providing computer
resources.

\end{document}